\documentclass{article}
\usepackage{spconf,amsmath,graphicx}
\usepackage{algorithm}
\usepackage[T1]{fontenc}
\usepackage{algorithmicx}
\usepackage{algpseudocode}
\usepackage{amsmath}
\usepackage{caption}
\usepackage{float}
\usepackage{pdfpages}
\usepackage{multirow}
\usepackage{indentfirst}
\usepackage{svg}
\usepackage{threeparttable}

\title{A Coarse-to-fine Morphological Approach With Knowledge-based Rules and Self-adapting Correction for Lung Nodules Segmentation}
\name{\begin{tabular}{c}Xinliang Fu$^{\star \dagger}$, Jiayin Zheng$^{\star \dagger}$, Juanyun Mai$^{\dagger}$, Yanbo Shao$^{\dagger}$, Minghao Wang$^{\dagger }$, \\
Linyu Li$^{\dagger}$, Zhaoqi Diao$^{\dagger}$, Yulong Chen$^{\S}$, Jianyu Xiao$^{\S}$, Jian You$^{\S}$, Airu Yin$^{\dagger \ddagger}$,\\Yang Yang$^{\pounds}$, Xiangcheng Qiu$^{\pounds}$, Jinsheng Tao$^{\pounds}$, Bo Wang$^{\pounds}$,
and Hua Ji$^{\pounds}$ $^{\dagger}$ $^{\ddagger}$
\thanks{ ${\star}$Having equal contribution with the first author
 ${\ddagger}$ Corresponding author}\end{tabular}}
\address{\begin{tabular}{c}${\dagger}$Advanced Medical Data Research Center, College of Computer Science, \\Nankai University, Tianjin, China\\
${\S}$ Department of \{Lung Cancer, Radiology\}, Tianjin Lung Cancer Center, \\Tianjin Medical University, Tianjin, China\\
${\pounds}$AnchorDx Medical Co., Guangzhou, China\\
Email:  \{1811112,1811523\}@mail.nankai.edu.cn, \{yinar, hua.ji\}@nankai.edu.cn\end{tabular}}

\begin{document}
%
\maketitle
\begin{abstract}
\noindent
The segmentation module which precisely outlines the nodules is a crucial step in a computer-aided diagnosis(CAD) system. The most challenging part of such a module is how to achieve high accuracy of the segmentation, especially for the juxtapleural, non-solid and small nodules. In this research, we present a coarse-to-fine methodology that greatly improves the thresholding method performance with a novel self-adapting correction algorithm and effectively removes noisy pixels with well-defined knowledge-based principles. Compared with recent strong morphological baselines, our algorithm, by combining dataset features, achieves state-of-the-art performance on both the public LIDC-IDRI dataset (DSC 0.699) and our private LC015 dataset (DSC 0.760) which closely approaches the SOTA deep learning-based models' performances. Furthermore, unlike most available morphological methods that can only segment the isolated and well-circumscribed nodules accurately, the precision of our method is totally independent of the nodule type or diameter, proving its applicability and generality.
\end{abstract}
\begin{figure*}
    \centering
    \includegraphics[scale=0.7]{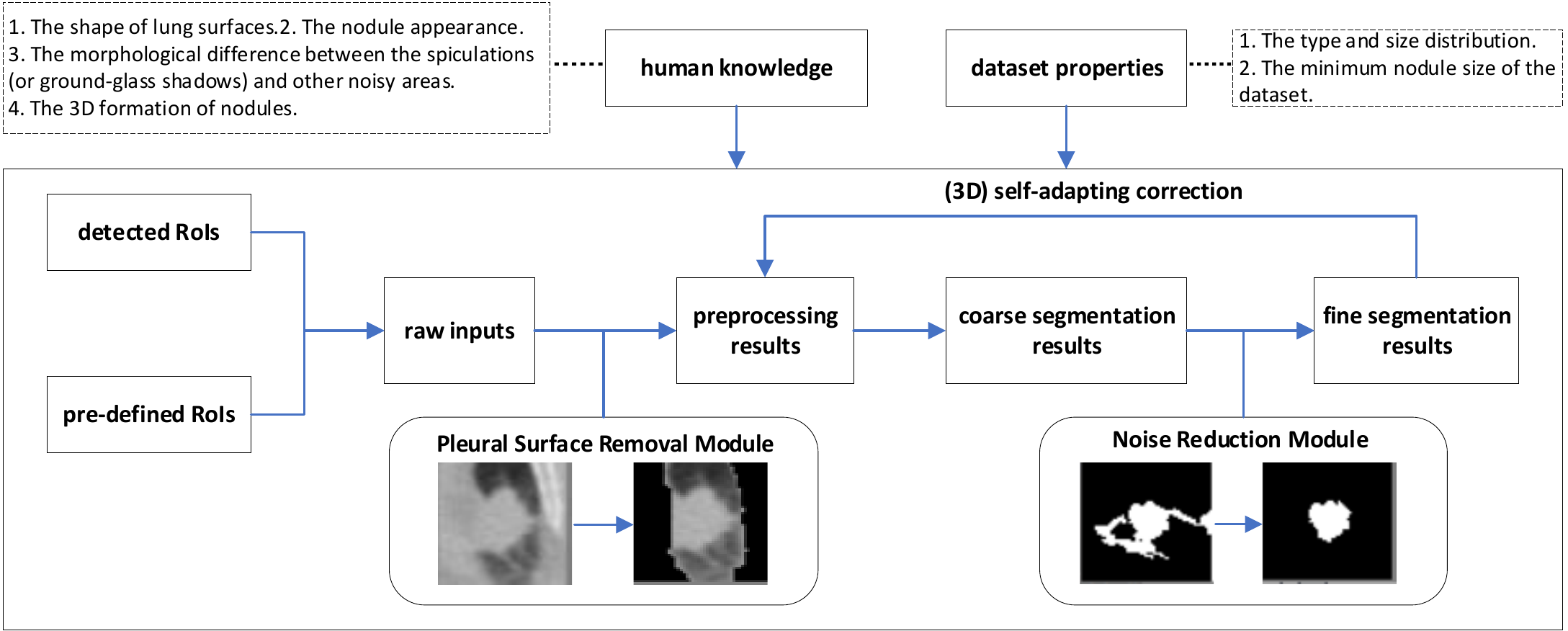}
    \captionsetup{font={footnotesize}}
    \caption{The overall pipeline of our coarse-to-fine framework. Taking the detection results or the pre-defined RoIs as inputs, our algorithm sequentially conducts lung wall removal, coarse segmentation and fine segmentation.  }
    \label{fig:my_label}
\end{figure*}
\section{Introduction}
\label{sec:intro}
\noindent
Pulmonary nodules are lung lesions with oval shape. Early diagnosis and investigation of these lung nodules are critical for enhancing patients' chances of survival and allowing effective treatments~\cite{1}. However, radiologists must scan hundreds of slices to distinguish between internal non-nodule structures and nodule cells, which takes time and effort. As a result, there is a high demand for an accurate and efficient computer-aided diagnostic (CAD) system~\cite{2}. The lung nodule segmentation module, which automatically outlines the nodule contours, is an important component of a CAD system~\cite{3}.

Many studies on automatic nodule segmentation have been conducted, and they can be classified into two groups: morphological methods and deep learning-based methods. The morphological methods including thresholding method~\cite{4,5}, region growing~\cite{6,7}, active contours~\cite{8,9}, and graph-cut methods~\cite{10,11}, are often based on pixel density; whereas deep learning-based methods~\cite{12,13} use neural networks to extract semantic information~\cite{14}. Although morphological techniques do not necessitate as much processing as DNN, their absence of deep semantic traits and 3D information might result in incorrect segmentation, particularly for juxtapleural~\cite{9}, ground-glass, and small-sized nodules~\cite{15}. To put it another way, their effects are size~\cite{5} or type dependent. On the other hand, deep learning models' interpretability and generalizability have been a significant barrier to their widespread adoption in healthcare settings. For instance, its performance is highly dependent on the designs and parameters of neural networks~\cite{16}; a model that performs well on one dataset may perform horribly on another one; there is no obvious rationale connecting the data about a case to the model's judgments~\cite{17}. Moreover, the interpretability is very critical for doctors to make clinical decisions. Hence, impediments to widely adoption of DL models in real healthcare settings remain strongly due to the black-box nature of DL. To this end, we propose a novel morphological algorithm to address the above issues. Our experimental results have shown that the algorithm can not only produce precise segmentations for nodules of various sorts but also provide excellent interpretability with detailed inference steps illustrations, making it well suited for a real-world CAD system. 

To summarize, our contributions are as follows:

1. As far as we know, our work is the first attempt to explicitly define knowledge-based principles extracted from clinical experience and knowledge to enhance the morphological method and conduct noise reduction. 

2. Our self-adapting thresholding method and its 3D-based version can produce better segmentation especially for small or ground-glass nodules, by referencing adjacent slices and gradually shrinking the bounding box. 

3. Our coarse-to-fine morphological method achieves state-of-the-art performance on both the LIDC-IDRI dataset (DSC 0.699) and LC015 dataset (DSC 0.760) and the effects are independent of the nodule type or size.

\section{Proposed Method}

\noindent
We start a coarse-to-fine segmentation method that removes the lung wall, selects candidate pixels, reduces noise, and applies self-adapting correction. As shown in Figure 1, our method does not necessarily need manually selected Region of Interests(RoIs) and can take the detection results as input, making it ideal for an end-to-end intelligent diagnosis system. 
\subsection{Knowledge-based Principles}
\noindent
\textbf{Rule 1}: 
An important property of convex figure is that every point on every line segment between two points inside or on the boundary of it remains inside or on the boundary, that is, S is a convex graphic if and only if for all two points p,q belong to S, the connectted line $\overline{pq}$ also belongs to S.

\noindent
\textbf{Rule 2}: If the bounding boxes circumstance the real nodules better, there will be less interference information.

\noindent
\textbf{Rule 3}: The ground-glass shadow pixels usually evenly surround the solid nodule structures and their mean pixel values is higher than that of the lung wall but lower than that of the solid nodule structure .

\noindent
\textbf{Rule 4}: The nodules are oval-shaped and spatially continuous; thus, compared to the original whole RoI box, the bounding box inherented from the more centered slice usually frames the contours of the topper slices more closely.

All the above principles are distilled from human knowledge or experience and will be utilized in the following steps. 

\subsection{Pleural Surface Removal}
\noindent
Outlining juxtapleural nodules precisely is difficult since the contours may contain the lung walls, reducing the thresholding method's performance. Several people used morphological methods~\cite{17} such as the global thresholding method~\cite{18} or deep neural networks~\cite{19} to segment the lung. However, using the entire CT image as input, these methods need a lot of processing and aren't good enough to eliminate all lung wall cells. Hence, we present a simple yet effective algorithm that needs only the RoI as input and can remove all pleural surface cells in the RoI.

\begin{figure}[h]
    \centering
    \includegraphics[scale=0.4]{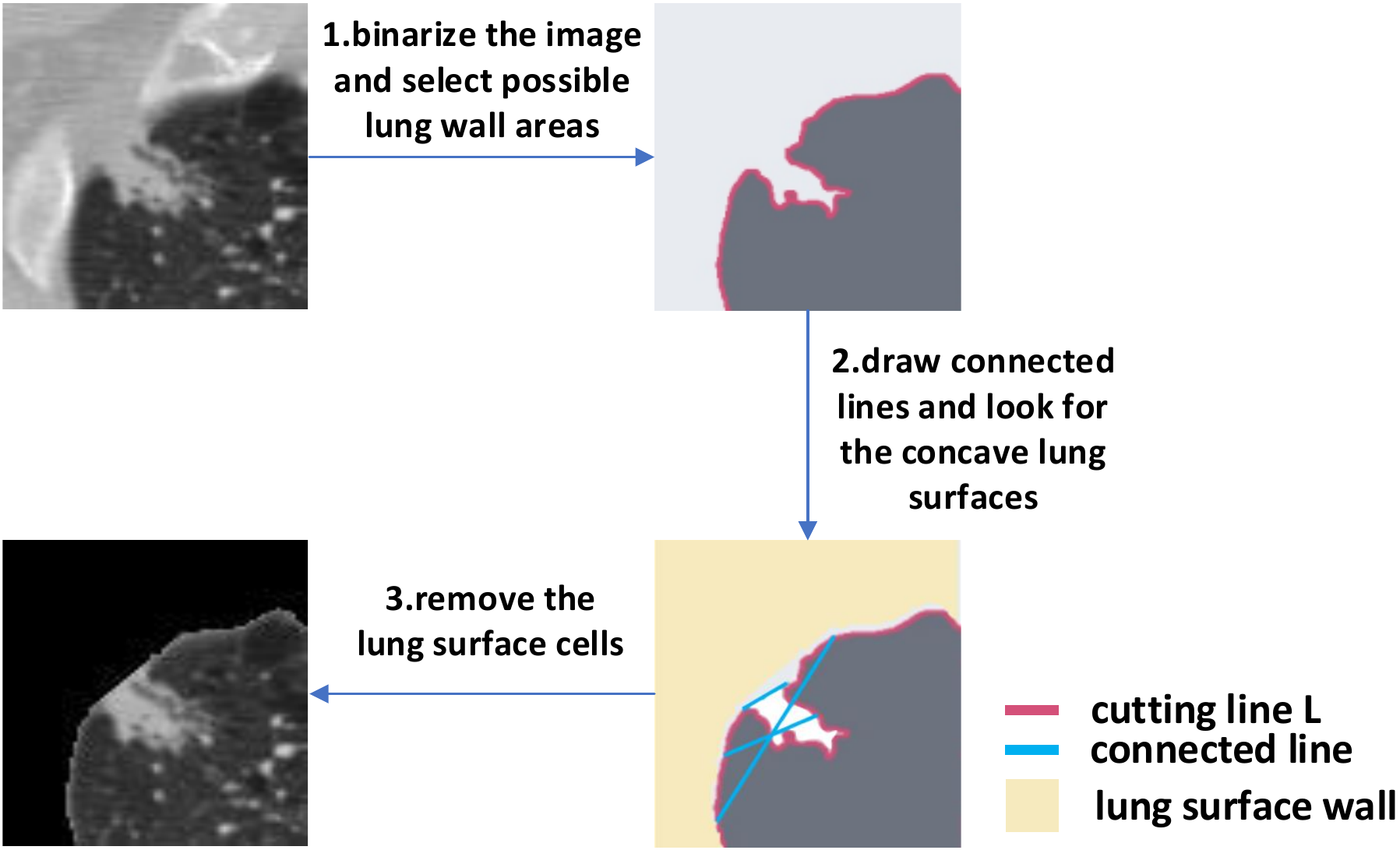}
    \captionsetup{font={footnotesize}}
    \caption{The grey area includes the concave lung wall area and the raised nodule structure. After drawing connected lines, we find that the yellow area is not covered by any line and remove the pixels in it. }
    \label{fig:my_label}
\end{figure}

To begin, we binarize the input image, changing the values of pixels above and below the average to 1 (shown grey in Figure 2) and 0 (shown black in Figure 2).  L stands for the cutting line that separates the grey and black areas. Since the white raised nodule and the black area together form a convex figure, according to the convex property mentioned in \textbf{Rule 1}, the line connecting any two points on L will only still in the white or black area, not crossing the lung surface wall area. Hence, we draw lines between any two spots on L, and the region that is not covered by any line is the lung surface area.

\subsection{Coarse Segmentation}
\noindent
There are two methods for coarse segmentation and they differ in how they select candidate pixels. The plain thresholding method treats pixels that are above average as candidate pixels, whereas the deformable method binarizes the image by Otsu`s method first and then does closing, subtraction and opening operations sequentially. Let  $s_m$ be the smallest nodule size in this dataset. There may still be multiple isolated areas after pixel selection, and we only keep those that are larger than $s_m$, referring to them as prospective nodule areas.

\subsection{Fine Segmentation}
\noindent
\subsubsection{Surrounding Noise Reduction}
\noindent
After selecting the nodule-like areas, we must remove the noisy pixels that surround the nodule structures, including the vasculature. To reduce noise, we need to find dividing lines that separate the nodule area from the noisy areas around it. The dividing line's start and end points should be on the boundary, and the line should be shorter than $\alpha$ pixels to avoid accidentally excluding part of the nodule. $\alpha$ is a hyper parameter that is set to 8 here. Furthermore, the noisy areas separated by the diving line should be greater than $\pi*(\dfrac{d+1}{2})^2$, where d is the length of the dividing line, to avoid improper removal of the nodular pixels near the boundary.
\subsubsection{Self-adapting Correction}
\noindent
The segmentation of small-sized nodule slices is challenging and the small nodules are divided into two types: 2D slices of nodules with small diameters (usually less than 5mm) and top or bottom slices of larger nodules. We concentrate on the latter in this article since nodules smaller than 5 mm are not the main focus of clinicians. 
If the bounding boxes circumstance the real nodules better, there will be less interference information (\textbf{Rule 2}). As a result, we propose a self-adapting correction method that shrinks the bounding boxes iteratively until they are close enough. Algorithm 1 describes the steps of the self-adapting correction. The iteration will stop if the current bounding box is close enough to frame the nodule contour. The way to judge whether or not the box is precise enough is by calculating the proportion of the nodule pixels. If the proportion is large enough, we can propose that there are not so many noisy pixels and thus stop the loop. 

However, the bounding box cannot be too small for two reasons. Firstly, to be a valid nodule outline, the contour outputted from the bounding box must be larger than $s_m$ where $s_m$ denotes the minimum nodule size. Secondly, for ground-glass nodules, a too-small box may exclude the ground-glass shadow by mistake and thus lead to inaccurate segmentation. The ground-glass shadows usually surround the solid nodule structures evenly (\textbf{Rule 3}). Hence, if the pixels of the difference between $box_{cur}$ and $box_{next}$ are equally distributed around the solid nodule structure, we propose that they belong to the ground-glass shadow and stop the iteration.  
\floatname{algorithm}{Algorithm}

\renewcommand{\algorithmicrequire}{input:}
\renewcommand{\algorithmicensure}{output:}
\begin{algorithm}[h]
    \footnotesize
    \captionsetup{font={footnotesize}}
    \caption{\textbf{Self-adapting Correction}}
    \begin{algorithmic}[1] 
        \Require \textbf{Raw input image}
        \Ensure  \textbf{The contours of a pulmonary nodule}
        \Function {Correct}{$original\_box,min\_nodule\_size,\epsilon$}
            \State $cur\_box \gets original\_box$
            \State $cur\_contour \gets$  \Call{threshold}{$cur\_box$}
            \State $next\_box \gets$ \Call{findbox}{$cur\_contour$}
            \While{$Area\_next\_box < Area\_cur\_box/\epsilon$}
            \State $next\_contour \gets$  \Call{threshold}{$next\_box$} 
            \State $cur\_box \gets next\_box$
            \State $next\_box \gets$ \Call{findbox}{$next\_contour$}
            \If{$next\_contour < min\_nodule\_size$}
                \State \Return{$cur\_contour$}
            \EndIf
            \State $cur\_contour \gets next\_contour$
            \EndWhile
            \State \Return{$next\_contour$}
        \EndFunction
    \end{algorithmic}
\end{algorithm}

\subsubsection{3D-based Self-adapting Correction}
\noindent
Inspired by the 3D formation of nodules (\textbf{Rule 4}) and how 3D neural network works, we offer a 3D-based self-adapting correction technique that outputs a closer bounding box based on the segments of nearby slices. Figure 3 depicts our method.
\begin{figure}[h]
    \includegraphics[scale=0.55]{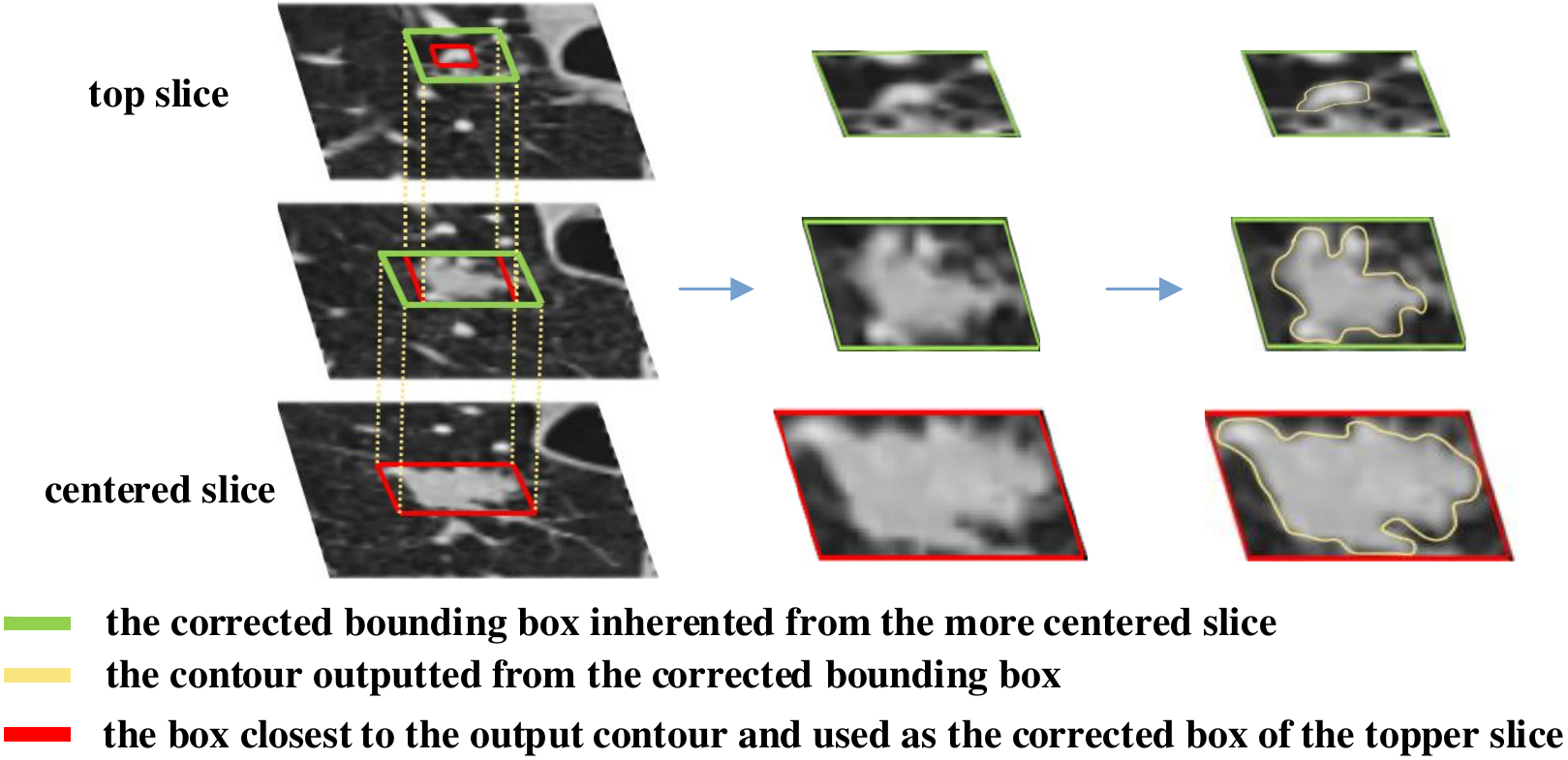}
    \captionsetup{font={footnotesize}}
    \caption{Example of the 3D-based self-adapting thresholding method. We utilize the segmentation result of the centered slice to help shrink the input bounding box for its adjacent slices and do this process iteratively until the top or bottom slices. }
    \label{fig:my_label}
\end{figure}
\begin{table*}[t]
\centering
\footnotesize
\begin{threeparttable}[b]
\caption[b]{Performance comparison between baseline approaches and ours under LIDC-IDRI and LC015 datasets.}
\begin{tabular}{|p{1.8cm}|l|p{0.3cm}p{0.3cm}p{0.4cm}p{0.4cm}p{0.3cm}p{0.4cm}p{1.1cm}|p{0.3cm}p{0.3cm}p{0.4cm}p{0.4cm}p{0.3cm}p{0.4cm}p{1.1cm}|}
\hline
\multicolumn{1}{|c|}{\multirow{2}{*}{Method}} & \multicolumn{1}{c|}{\multirow{2}{*}{Input}}  & \multicolumn{7}{c|}{LIDC-IDRI Dataset}  & \multicolumn{7}{c|}{LC015 Dataset}\\ \cline{3-16} 
\multicolumn{1}{|c|}{}                        & \multicolumn{1}{c|}{}                                       & Avg.      & Solid  & mGGN & pGGN    & (0,10)   & [10,20)  &  [20,inf){\scriptsize mm}  &  Avg.      & Solid  & mGGN & pGGN     & (0,10)   & [10,20)  &  [20,inf){\scriptsize mm}   \\ \hline
FMM(MM)~\cite{20}                                          & \begin{tabular}[c]{@{}l@{}}pre-defined RoIs\end{tabular} & 0.396   & 0.427 & 0.271        & 0.179     & 0.309       & 0.504        & 0.468       & 0.376   & 0.518 & 0.399        & 0.177     & 0.358          & 0.369          & 0.389          \\ \hline
PTM(MM)                     & \begin{tabular}[c]{@{}l@{}}pre-defined RoIs\end{tabular} & 0.601   & 0.604 & 0.575        & 0.582     & 0.527       & 0.662        & 0.692       & 0.720   & 0.732 & 0.721        & 0.704     & 0.651          & 0.721          & 0.736          \\ \hline
PDM(MM)                                & \begin{tabular}[c]{@{}l@{}}pre-defined RoIs\end{tabular} & 0.673   & 0.686 & 0.627        & 0.581     & 0.642       & 0.702        & 0.720       & 0.678   & 0.719 & 0.682        & 0.629     & 0.631          & 0.675          & 0.693          \\ \hline
Ours(MM)                                           & \begin{tabular}[c]{@{}l@{}}pre-defined  RoIs\end{tabular} & \textbf{0.699}   & \textbf{0.709} & \textbf{0.646}        & \textbf{0.640}     & \textbf{0.683}       & \textbf{0.719}        & \textbf{0.712}       & \textbf{0.760}   & \textbf{0.770} & \textbf{0.760}        & \textbf{0.752}     & \textbf{0.737} & \textbf{0.760} & \textbf{0.766} \\ \hline
Ours(MM)                                           & \begin{tabular}[c]{@{}l@{}}detected  RoIs\end{tabular}   &         &       &              &           &             &              &             & 0.682   & 0.701 & 0.667        & 0.727     & 0.704          & 0.700          & 0.657          \\ \hline
nnUNet(DL)~\cite{16}                                      & whole image                                                 &\textbf{0.818}         &   \textbf{0.826}    &       \textbf{0.776}       &    \textbf{0.733}       &     \textbf{0.799}        &      \textbf{0.837}        &      \textbf{0.863}       &   \textbf{0.781}      &   \textbf{0.823}    &      \textbf{0.778}        &      \textbf{0.762}     &      \textbf{0.741}          &        \textbf{0.792}        &       \textbf{0.798}         \\ \hline
\end{tabular}
\begin{tablenotes}
\item[1] MM:morphological methods $^2$ DL:deep learning $^3$ PTM:Plain Thresholding Method $^4$ PDM:Plain Deformable Method $^5$ mGGN:mix ground-glass nodule $^6$ pGGN:pure ground-glass nodule 
\end{tablenotes}
\end{threeparttable}
\end{table*}
\section{Experiment}
\label{sec:typestyle}

\subsection{Dataset}

\noindent
 We test the performance of our method on the public LIDC-IDRI dataset~\cite{21} and our private LC015 dataset. The LIDC-IDRI dataset provides labeled nodule outlines for 1018 patients (with 2651 nodules), while LC015 dataset is gathered from 14 hospitals and includes 990 patients (with 1186 nodules). In two ways, our private LC015 dataset differs from the present public datasets. First, the nodules are confirmed by clinicians rather than simply labeled  by radiologists as such in public datasets by radiologists. Furthermore, the LC015 dataset has a more equally distributed data set than the LIDC-IDRI dataset, in which most nodules are tiny or solid. Therefore, we can see how different techniques function with varied nodule types and sizes by experimenting on two datasets.

\begin{figure}[h]
    \includegraphics[scale=0.25]{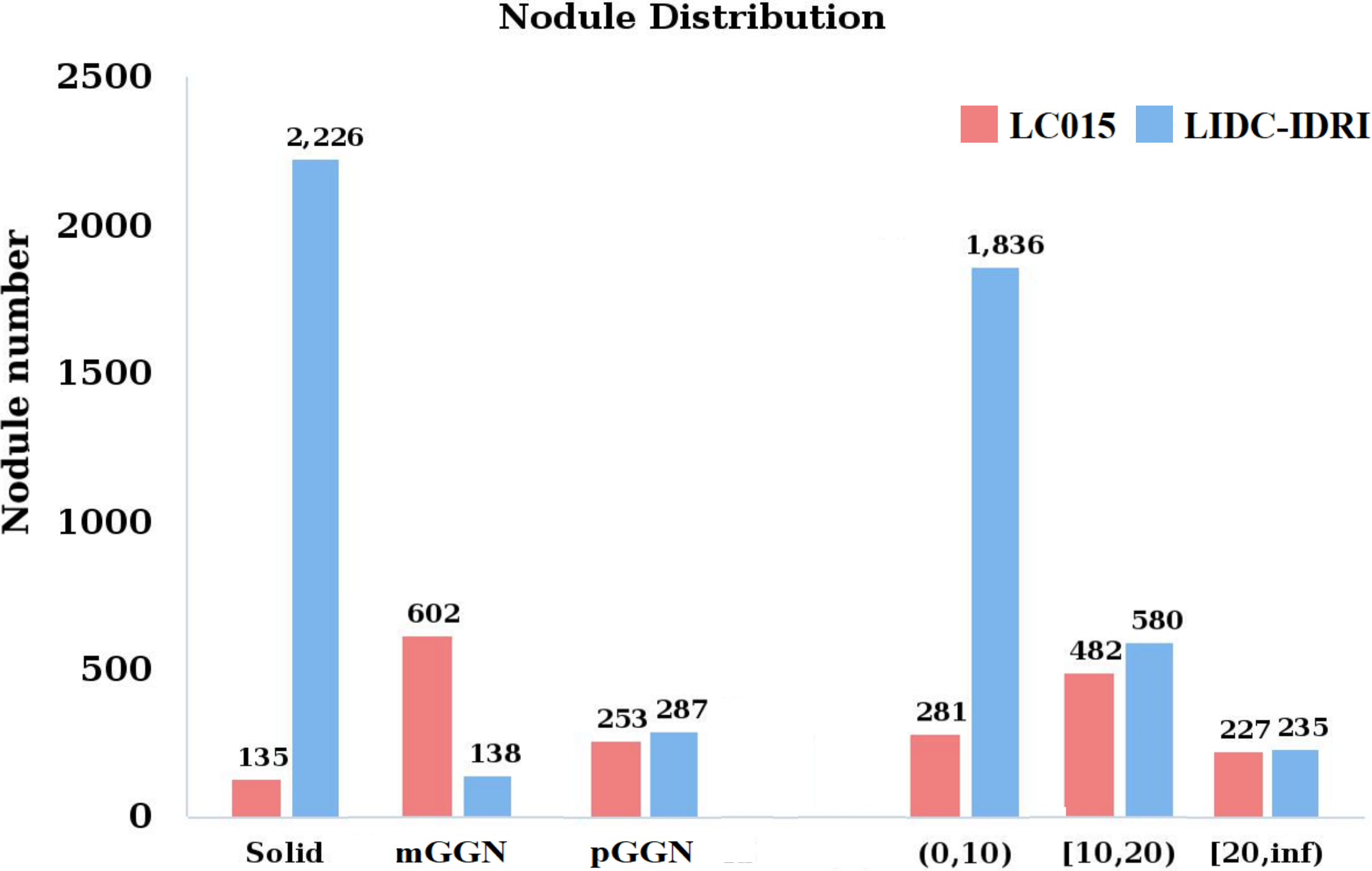}
    \captionsetup{font={footnotesize}}
    \caption{Nodule size and type distribution of LIDC-IDRI and LC015 datasets.}
    \label{fig:my_label}
\end{figure}
\subsection{Evaluation Metrics} 
\noindent
The Dice Similarity Coefficient (DSC) is a commonly used metric for determining the degree of overlap between the predicted and ground-truth segments.The higher the value is, the more precise the segmentation is~\cite{22,23}.
\subsection{Performance Comparison}
\noindent
Unlike previous morphological works that do evaluations only on a subset of the dataset~\cite{20}, we conduct experiments on the whole dataset (3837 nodules in total).

When compared to SOTA morphological approaches, plain thresholding, and deformable methods, our proposed approach produces the best segmentation results on both datasets, as shown in Table 1. When employing pre-defined RoIs, our morphological method closely approaches the nnUNet model on the LC015 dataset. This demonstrates that the morphological method can reach remarkable performance with a faster speed and a clearer rule-based illustration.

Furthermore, earlier morphological approaches typically struggle with juxtapleural, non-solid, and tiny nodules. On the LIDC-IDRI dataset, for example, FFM's performance on non-solid and small nodules drops by 54.9\% and 22.0\%, respectively. By contrast, our method has consistently high performance that is unaffected by nodule type or diameter. Moreover, even when using the detected RoIs as input where more noisy pixels are introduced, our algorithm can still perform well and output precise segmentations. The results above fully demonstrate the superiority and necessity of incorporating knowledge-based rules and self-adapting correction. 

\section{Conclusions}
\noindent

By seamlessly integrating knowledge-based principles and a well designed self-adapting correction algorithm, we present an efficient image processing framework that can yield exact segments for nodules of all sorts and meanwhile provide good explainability of the underlying's inference steps. Compared to other related work, our framework provides two outstanding scientific merits: First, the rule-based and self-correction methods well support accurately segmenting previously intractable nodules including juxtapleural, non-solid, and small nodules. Second, our methods' great precision proves that morphological algorithm can achieve the same accuracy as DL models while using less processing power and meanwhile having much better interpretability and explainability. Therefore, we claim that our approach is well suited for a real-world CAD system that requires precision, automation, and explanation for clinical adoption.

\bibliographystyle{IEEEbib}
\bibliography{main.bib}

\end{document}